\documentclass[10pt, conference, letterpaper]{IEEEtran}

\def\BibTeX{{\rm B\kern-.05em{\sc i\kern-.025em b}\kern-.08emT\kern-.1667em\lower.7ex\hbox{E}\kern-.125emX}}

\usepackage{tikz}
\usepackage{amsmath,amssymb}
\usepackage{filecontents}

\usepackage[english]{babel}

\newcommand{\var}[1]{\textbf{#1}}
\usepackage{array}
\newcolumntype{H}{>{\setbox0=\hbox\bgroup}c<{\egroup}@{}}

\usepackage{hyperref}
\usepackage{algorithmic}
\usepackage[nointegrals]{wasysym}
\usepackage{graphicx}
\usepackage{textcomp}
\usepackage{xcolor}
\usepackage{colortbl}
\usepackage{underscore}
\usepackage{multirow}
\usepackage{array}
\usepackage{balance}
\usepackage{tabularx}
\usepackage{booktabs}
\usepackage{amsmath}
\usepackage{mathtools}
\usepackage{float}

\usepackage{algorithmic}
\usepackage[normalem]{ulem}
\usepackage{comment}
\usepackage{subfigure}

\usepackage{fancyhdr}
\usepackage{extramarks}

\begin{document}

\date{}

\title{\Large \bf Postcertificates for Revocation Transparency}

\author{
{\rm Nikita Korzhitskii}\\
Link\"oping University
\and
{\rm Matus Nemec}\\
Link\"oping University
\and
{\rm Niklas Carlsson}\\
Link\"oping University
}

\pagestyle{plain}
\maketitle

\begin{abstract}
The modern Internet is highly dependent on trust communicated via certificates. However, in some cases, certificates become untrusted, and it is necessary to revoke them. In practice, the problem of secure revocation is still open. Furthermore, the existing procedures do not leave a transparent and immutable revocation history. We propose and evaluate a new revocation transparency protocol that introduces postcertificates and utilizes the existing Certificate Transparency (CT) logs. The protocol is practical, has a low deployment cost, provides an immutable history of revocations, enables delegation, and helps to detect revocation-related misbehavior by certificate authorities (CAs). With this protocol, a holder of a postcertificate can bypass the issuing CA and autonomously initiate the revocation process via submission of the postcertificate to a CT log. The CAs are required to monitor CT logs and proceed with the revocation upon detection of a postcertificate. Revocation status delivery is performed independently and with an arbitrary status protocol. Postcertificates can increase the accountability of the CAs and empower the certificate owners by giving them additional control over the status of the certificates. We evaluate the protocol, measure log and monitor performance, and conclude that it is possible to provide revocation transparency using existing CT logs. 
\end{abstract}

\section{Introduction}

\begin{figure*}[t]
  \centering
  \subfigure[Standard CT submission process: A precertificate 
  (identical to the certificate-to-be-issued 
  except for a poison extension)
  is submitted.]{
    \includegraphics[trim = 00mm 060mm 340mm 000mm,clip, width=0.22\textwidth]{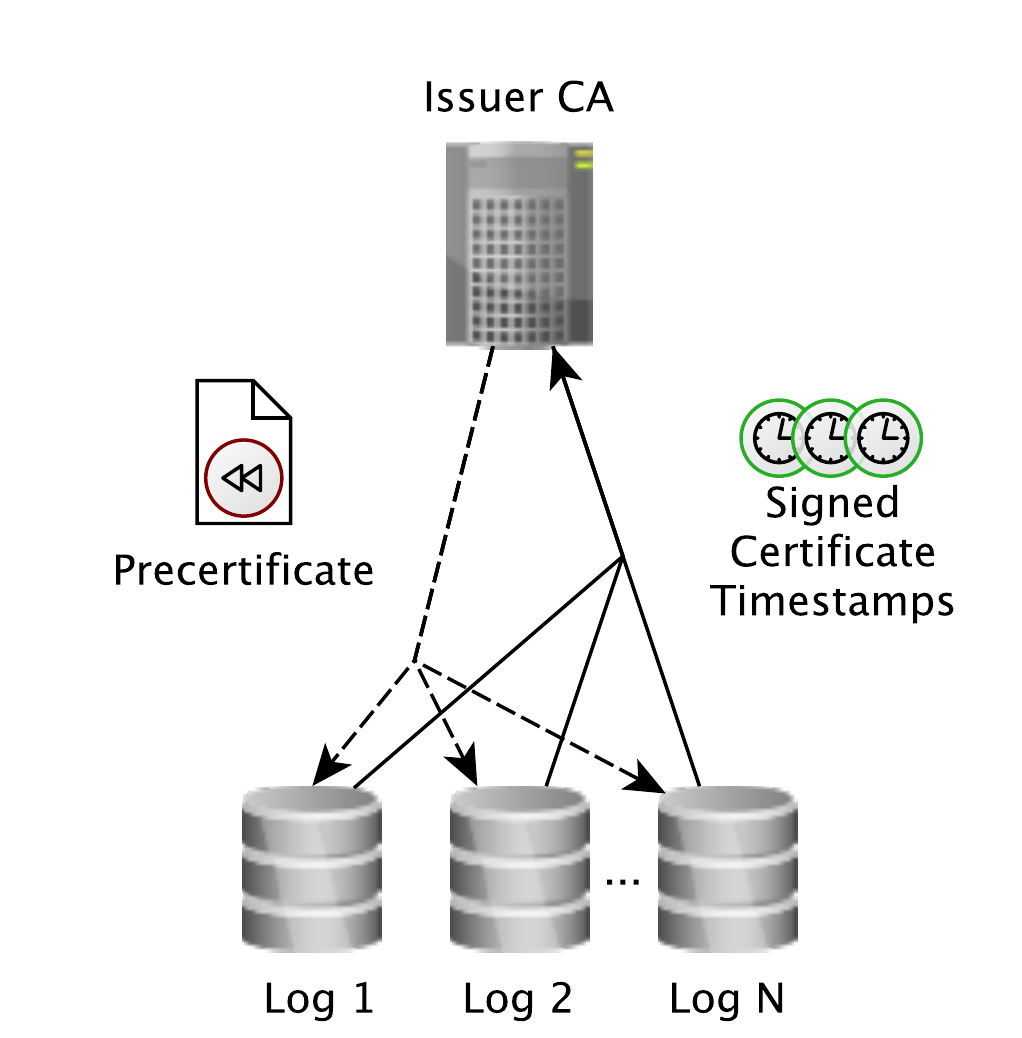}}
  \hspace{6pt}
   \subfigure[Issuance of a certificate with 
   SCTs.
   Postcertificates can be issued and delivered to the client at this stage.]{
    \includegraphics[trim = 120mm 61mm 230mm 0mm,clip, width=0.19\textwidth]{figures/postcertificates-graphic-general2.pdf}}
    \hspace{6pt}
   \subfigure[Transparent revocation by the client. The client can bypass the issuer CA and post a postcertificate to logs of other operators. CAs are required to monitor the logs and proceed with the revocation upon the detection of a postcertificate.]{
    \includegraphics[trim = 236mm 63mm 0mm 22mm,clip, width=0.49\textwidth]{figures/postcertificates-graphic-general2.pdf}}
    \vspace{-6pt}
  \caption{Certificate Transparency and Revocation Transparency with postcertificates.}
  \label{fig:time-revoked}
      \vspace{-6pt}
\end{figure*}

The Web Public Key Infrastructure (WebPKI) is widely used to establish trust on the Internet. It consists of many trusted third-parties, so-called certificate authorities (CAs), that, among other things, associate public keys to Internet names by issuing cryptographically signed certificates.
Most existing CAs, irrespectively of their size, are capable of issuing certificates for any Internet name. Due to previous incidents~\cite{amann2017mission,zhang2014analysis} and the need for tight oversight, Google proposed and enforced Certificate Transparency (CT)~\cite{RFCCT,scheitle2018rise}. 
CT is a protocol that facilitates the detection of misissuance via reliable logging of all WebPKI certificates.

According to 
current
CA/Browser conventions~\cite{CABR20}, issuers must support certificate revocation and provide certificate status via Online Certificate Status Protocol (OCSP)~\cite{RFCOCSP}. Revocation is needed 
when a certificate must be invalidated, e.g., due to private key loss, security breaches, and other issues that render certificates insecure. At all times, CAs must provide up-to-date revocation statuses for all 
valid certificates~\cite{CABR20}. 

The adopted revocation protocols have many security problems~\cite{chuat2019sok,liu2015end,zhang2014analysis}. Moreover, most browsers do not perform full revocation status checks for every certificate~\cite{chuat2019sok}. Instead, some browsers use proprietary revocation protocols~\cite{CRLSets,crlite}. In addition, most revocation statuses of certificates disappear soon after certificate expiration~\cite{korzhitskii2021revocation}. This status-handling practice motivates the deployment of a revocation transparency protocol that would preserve revocation history. For a client, certificate revocation is often a lengthy (and expensive) process that can only be performed through the issuing CA. The above drawback highlights the need for autonomous revocation, i.e., clients should be able to control the validity of their certificates independently of the issuing CA~\cite{chuat2019sok}. 

Revocation Transparency~\cite{laurie2012revocation} was originally proposed by Google as a mechanism for storage and dissemination of revocation statuses. Many other protocols and modifications to WebPKI have been proposed to provide revocation transparency (see Section~\ref{section:relatedwork}). However, most of these protocols require substantial changes or a complete replacement of the infrastructure.

Currently, no revocation transparency protocol has been adopted.
Hence, the WebPKI lacks autonomous revocation as well as a transparent and immutable history of all revocations.
The absence of such history makes certificate revocations difficult to study. 
Nowadays, it is necessary to regularly perform large-scale OCSP measurements.
Moreover, lack of transparency complicates the detection of revocation-related misbehavior by CAs, e.g., advertisement of incorrect or contradictory revocation statuses. 
A revocation transparency protocol that preserves revocation history will provide a valuable record of revocation-related misbehavior, mass-revocation events, and WebPKI revocation practices.

In this paper, we introduce a practical and incrementally-deployable protocol for logging special postcertificates in existing certificate transparency logs. The protocol is more easily deployable than the previously proposed revocation status and revocation transparency protocols. Postcertificates enable {\it autonomous revocation by policy}, revocation delegation, and transparent preservation of revocation requests.
Furthermore, the protocol increases the accountability of the CAs and empowers the certificate owners by giving them additional control over the status of the certificates.

Outline:
In Section~\ref{section:postcertificate} we describe the postcertificates and evaluate our proposal against an existing 
framework for evaluation of 
revocation and delegation methods. To demonstrate that existing CT logs can support the deployment of postcertificates, we present performance measurements of existing CT logs in Section~\ref{section:performance}.
Next, in Section~\ref{section:deployability}, we analyze the deployability of postcertificates and model the potential impact on the logs. 
Section~\ref{section:relatedwork} overviews related work and Section~\ref{section:conclusion} presents the conclusion. 

\section{Postcertificate}
\label{section:postcertificate}
In this paper, we introduce the {\it postcertificates} and show that they can be used together with the existing CT logs to improve the transparency of revocations.

The concept of postcertificates requires the notion of {\it precertificates} from the current CT standard~\cite{RFCCT}.
Precertificates are identical to certificates that are going to be issued, with the exception that they contain a critical poison 
extension. 
The poison extension renders the precertificates invalid in browsers.
Precertificates are created and logged prior to certificate issuance 
to obtain inclusion promises --- Signed Certificate Timestamps (SCTs). These SCTs are later embedded into the issued certificates or delivered using TLS or OCSP stapling. 

Similarly, a {\it postcertificate} corresponds to a certificate-to-be-revoked, and it contains 
a critical poison extension that renders the postcertificate invalid in browsers.\footnote{An original certificate, a precertificate, and a proposed postcertificate share most of the data through TBSCertificate certificate field~\cite{RFCPKI}.}

\subsection{Revocation requests via postcertificates}

We propose that every revocation should start with the submission of a postcertificate to a CT log. 
For a new submission, a CT log issues an SCT that contains the timestamp of the submission. The SCT is returned to the submitter directly, while the submitted entry and the corresponding timestamp are guaranteed to be published in the log within some $MaxMergeDelay$ starting from the submission\footnote{The SCTs could be used to prove CT log misbehavior and to reduce revocation delay. See Sections~\ref{sub:ca-misbehavior} and \ref{sub:revocation-delay}.}.
A postcertificate and proof of its inclusion in a trusted log verifiably constitute a revocation request.
For postcertificate revocation to function, {\it CAs must regularly monitor} their own and other trusted CT logs, discover postcertificates, and update the revocation statuses according to the
found postcertificates.

A postcertificate together with the earliest issued SCT constitute a timestamped revocation request, but not the status itself, whereas the statuses are advertised by an underlying revocation status protocol. This requires a log insertion operation only for an actual status change and not for every regularly issued status\footnote{Statuses are frequently reissued~\cite{CABR20}, while the actual status of a certificate (revoked/non-revoked) remains the same most of the time. See Section~\ref{sub:status-update}.}. After discovering a postcertificate, a CA must deliver an updated, authenticated, and timestamped revocation status to the clients. The status update must be performed using the underlying methods specified in the revoked certificate (e.g., CRL~\cite{RFCPKI}, OCSP~\cite{RFCOCSP}) within some $MaxRevocationDelay$. The delay can be defined to start from the submission time or publication time of a postcertificate. The protocol is illustrated in Figure~\ref{fig:time-revoked}.

Postcertificates are structurally identical to X.509 certificates; hence, the time complexity of inclusion, lookup, and proof operations in CT logs is not affected.

\subsection{Proving CA misbehavior}
\label{sub:ca-misbehavior}
Currently, CT monitors observe certificate issuance.
In the proposed protocol, CAs and other third parties such as revocation monitors, software vendors, and log operators similarly track the contents of the logs and disclose revocation misbehavior.
Upon discovery, third parties should provide inconsistent revocation status responses to the public for verification. Other third parties can verify the disclosures and proceed with the appropriate action.

\begin{figure}[t]
  \centering
  \subfigure[Normal revocation]{
    \includegraphics[trim = 32mm 99mm 25mm 19mm,clip, width=0.4\textwidth]{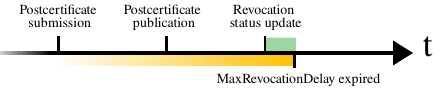}}\\
  \vspace{-6pt}
  \subfigure[Late or incorrect revocation status update after the expiration of Maximum Revocation Delay]{
    \includegraphics[trim = 32mm 71mm 25mm 47mm,clip, width=0.4\textwidth]{figures/arrows6.pdf}}\\
  \vspace{-6pt}
   \subfigure[Early revocation without postcertificate submission]{
     \includegraphics[trim = 36.5mm 45mm 33mm 70mm,clip, width=0.22\textwidth]{figures/arrows6.pdf}}
    \hspace{6pt}
   \subfigure[Revocation without a postcertificate submission]{
       \includegraphics[trim = 37mm 20mm 33mm 95mm,clip, width=0.22\textwidth]{figures/arrows6.pdf}}
    \hspace{6pt}
    \vspace{-8pt}
  \caption{Revocation via postcertificates and types of revocation misbehavior by CAs. $MaxRevocationDelay$ can be standardized to start from either postcertificate submission time or postcertificate publication time.}
  \label{fig:misbehavior}
      \vspace{-6pt}
\end{figure}

\begin{table*}[t]
\centering
\caption{Cases of CA revocation misbehavior with earliest proof times, requirements, and verification steps.  
We evaluate two definitions of $MRD$ \{(A) from submission, and (B) from publication\} and three CA misbehavior cases \{({\bf M1}) a missing status update within the expiration of some $MRD$, ({\bf M2}) delivery of an incorrect revocation status after the $MRD$, and ({\bf M3}) delivery of a ``non-good'' status before a postcertificate has been submitted to a trusted CT log\}. 
}
\vspace{-6pt}
\label{tab:cases-misbehavior}
\resizebox{0.99\textwidth}{!}{%

\begin{tabular}{llp{3.2cm}p{6cm}p{6.5cm}}
\hline
\multicolumn{1}{c}{}	&	{\bf Case}	&	{\bf Earliest proof time ($t_{proof}$)}	&	{\bf Misbehavior proof }	&	{\bf Verification } \\
\hline
\multirow[c]{2}{23mm}{MRD starts from submission, \newline $MRD_A > MMD$}	& {\bf M1}, {\bf M2}	&

$ E [ t_{submission}] + MRD_{A}$ & 

Published entry $E$, $STH$  
\newline Incorrect status $S$ 
\newline $ A := MerkleAuditProof(E, STH[ treesize ])$ 
&	

Verify $ E[P], STH, S, A$. 
\newline $S[ status]  \neq (E[ P])[ status] $ 
\newline $S[ t ]  \ge t_{proof}$, $S[ C]  = (E[ P]) [ C] $ 
\newline   $E [ L ] = STH [ L ]$ 
\newline $E[ number ] < STH[ treesize ]$  \\

\cline{2-5}
	&	{\bf M3}	&	$S [ t ] + MMD$	&	
	
	Early status $S$ \newline 
	Set $Q$ of single $STH$s for each $L \in {\cal L}$ \newline
	For each $STH \in Q$, $STH[t] \geq t_{proof}$
	&
	$S[status]  \neq ``good"$. Verify $S$.  \newline
	For each $STH \in Q$, 
	verify $STH$,
	fetch all $E$ preceding the $STH$ from log $STH[L]$. 
	
	
	Verify that $\nexists E | \big((E[P])[C] = S[C] \wedge E[t] < S[t]\big) $. \\
	
	\hline
\multirow{2}{23mm}{MRD starts from publication, \newline $MRD_B > 0$} &
{\bf M1}, {\bf M2}	&

$STH[t]$$+$$MRD_{B}$, where $STH$ is the earliest detected STH covering the submitted postcertificate $P$  & 

Same as cases {\bf M1}, {\bf M2} with $MRD_A$ 
&

Same as cases {\bf M1}, {\bf M2} with $MRD_A$\\

	\cline{2-5}
	&	{\bf M3}	
	&	$S[t] + MMD$ &	
Same as case {\bf M3} with $MRD_A$ 
&
Same as case {\bf M3} with $MRD_A$\\
\end{tabular}}
\vspace{-6pt}
\end{table*}

Revocation misbehavior by a CA may be defined as ({\bf M1}) a missing status update within the expiration of some Maximum Revocation Delay, ({\bf M2}) delivery of an incorrect revocation status after the Maximum Revocation Delay, and ({\bf M3}) delivery of a ``non-good'' status before a postcertificate has been submitted to a trusted CT log. Normal revocation using a postcertificate is illustrated in Figure~\ref{fig:misbehavior}(a).
Cases of misbehavior ({\bf M1, M2}) are illustrated in Figure~\ref{fig:misbehavior}(b), and ({\bf M3})
is illustrated in Figures~\ref{fig:misbehavior}(c) and \ref{fig:misbehavior}(d).

The cases of misbehavior can be proven and verified only after time $t_{proof}$, which depends on the way $MaxRevocationDelay$ is defined. The delay can start either at the submission time or at the publication time of a postcertificate. The latter potentially enables faster misbehavior proof time in a real-life scenario, when submission-to-publication delays of logs are low.
Section~\ref{section:performance} presents a measurement of the current submission-to-publication delays of logs.

Table~\ref{tab:cases-misbehavior} summarizes the types of misbehavior and provides the earliest times ($t_{proof}$) when the three cases of CA misbehavior ({\bf M1, M2, M3}) can be detected and proved. Table~\ref{tab:cases-misbehavior} also provides proof and verification requirements that a third party must satisfy to claim CA misbehavior.
The notation is summarized in Table~\ref{tab:my_label}.

In general, to prove {\bf M1} and {\bf M2}, a third party must provide a published postcertificate entry, proof of its inclusion, a recent STH, and an incorrect status. However, to prove {\bf M3} a third party must verify that no trusted log contains a postcertificate that was submitted before an early status was published.

Currently, most instances of CA misbehavior are first publicly announced 
using 
a thread in the CT-policy Google 
group (e.g.,~\cite{ChromeFullCT}).
By analogy with the CT standard~\cite{RFCCT}, we do not explicitly specify how or when monitors/clients/third-parties should monitor CT logs for postcertificates. 
However, similarly to the monitoring of certificate issuance, software
vendors can implement postcertificate monitoring as part of browsers, extensions, standalone, or distributed applications. Monitoring could also be crowd-sourced via 
correspondence checks of the published statuses and published postcertificates, triggered by an outdated OCSP response, a soft or hard-fail, 
etc.

CAs must only provide a revoking status for a certificate after submitting the corresponding postcertificate to the logs. Hence, CAs and logs must synchronize their clocks, as the timestamps of the postcertificate revocation requests $E[t]$ and status time $S[t]$ now depend on the clocks of the CT logs. In Section~\ref{subsection:sync} we take a closer look at the current clock synchronization of the existing CT logs.

It is necessary to submit postcertificates to several independently-operated logs because a misbehaving log (e.g., one that belongs to the CA that issued a postcertificate) can reject a postcertificate submission, or ``forget"~\cite{dowling2016secure} to include a postcertificate.  In the latter case, the postcertificate submitter can disclose the SCTs obtained during the submission and prove log misbehavior.

\begin{table}[t]
    \centering
    \vspace{-6pt}
    \caption{Description of notation}
    \label{tab:my_label}
    \vspace{-6pt}
    \resizebox{0.49\textwidth}{!}{%
        \begin{tabular}{ll}
            \hline
            {\bf Notation} & {\bf Description} \\
            \hline
            $E(P, t_{submission}, L, number)$ & Published numbered entry of $P$ in a log $L$ \\
            $MerkleAuditProof(E, treesize)$ & Merkle Audit proof for $E$ relative to $treesize$ \\
            $MMD$, $MaxMergeDelay$ & Maximum Merge Delay \\
            $MRD$, $MaxRevocationDelay$ & Maximum Revocation Delay \\
            $P(C, status)$ & Postcertificate with $status$ for certificate $C$ \\
            $S(C, status, t)$ & Revocation $status$ of certificate $C$ at time $t$ \\ 
            $STH(L, t, treesize, hash, sig)$ & Signed Tree Head of log $L$ signed at time $t$ \\
            $t_{proof}$ & Earliest possible CA misbehavior proof time \\
            ${\cal L}$ & Set of all trusted logs\\
        \end{tabular}
}
    \vspace{-6pt}
\end{table}

\subsection{Postcertificate schemes}

We propose two alternative postcertificate schemes: CA-issued and self-signed. The evaluation of the schemes is presented in Table~\ref{tab:comparison}. We omit the comparison between postcertificates and the original Revocation Transparency proposal~\cite{laurie2012revocation}. The original proposal introduces new logs that act as status endpoints, i.e., the proposal intends to replace other revocation status protocols. Instead, postcertificates intend to complement the existing and future revocation status protocols.

{\bf CA-issued postcertificate:}
After issuing a regular certificate, a CA can issue a copy of a target certificate with a critical X.509 revocation extension --- a postcertificate. The extension prevents the postcertificate from being correctly validated as a target certificate (by analogy with precertificates). This postcertificate could then be shared with the owner of the private key. Any party in possession of a postcertificate can at any time initiate revocation of the corresponding certificate without directly contacting the CA by simply submitting the postcertificate to a trusted log (e.g., from Apple's \cite{AppleCTLogProgram} or Google's trusted log list \cite{LogsChrome}). 

{\bf Self-signed postcertificate:}
A self-signed postcertificate is also a copy of a target certificate with a critical X.509 revocation extension. However, in this case, the postcertificate is issued
by a holder of the private key rather than the CA. 
Self-signed postcertificates somewhat resemble {\it proxy certificates}~\cite{chuat2019sok,welch2004x} since they are delegated to revoke their parent certificates. In this case, a self-signed postcertificate cannot be considered a valid X.509 certificate since (a) the specified issuer does not correspond to the real one and (b) the certificate is signed with the private key of the certificate-to-be-revoked. Note that for self-signed postcertificate revocation to work, CT logs must modify their certificate and chain validation processes, while {\it CA-issued postcertificates require no changes to the logs}. CT logs have to accept self-signed postcertificates with the certificate chains extended with the certificates that are being revoked.
In this scheme, the private-key owners can issue self-signed postcertificates at any time and initiate revocations through third-party logs.

{\bf Revocation extension:}
Both postcertificate types include a revocation extension similar to the poison extension from the CT standard~\cite{RFCCT}. 
Here, we describe some of additional functions that the revocation extension may provide. By default, the revocation extension designates a certificate as a postcertificate and renders it invalid in browsers.
However, for both types of postcertificates, a revocation extension may provide additional information such as revocation reason and invalidation date.
If the ``CA-issued postcertificate'' scheme is adopted, CAs may be required to issue several postcertificates (e.g., with different revocation reasons) upon request. That would allow a certificate owner to revoke the certificate and specify additional parameters through the submission of the most relevant postcertificate to CT logs.
Clients may request postcertificates with different invalidation dates and reason codes, which would allow the clients to precisely control the revocation statuses. By default, the invalidation date of a revocation status would be equal to the postcertificate submission time, and the revocation reason is not specified.

\subsection{Evaluation of postcertificate schemes using framework~\cite{chuat2019sok}}

\begin{table*}[t]
\centering
\caption{Evaluation of postcertificates}
\label{tab:comparison}
\vspace{-6pt}
\resizebox{0.8\textwidth}{!}{%

\begin{tabular}{@{}llllH@{}}

\toprule

 & \textbf{Property} & \textbf{CA-issued postcertificate} & \textbf{Self-signed postcertificate} & \textbf{Revocation Transparency \cite{laurie2012revocation}} \\ 

\midrule

& \textbf{Postcertificate issued by} & CA & Private key holder & --- (no additional certificates) \\
& \textbf{Postcertificate chains back to} & CA certificate & Original certificate & --- (no additional certificates) \\
& \textbf{Parties capable of revocation} & Postcertificate holder & Postcertificate holder, private key holder & Not specified \\
& \textbf{CA-bypassing revocation} & \CIRCLE \space(iff a postcertificate was shared) & \CIRCLE & Not specified \\
& \textbf{CAs must monitor CT logs} & \CIRCLE & \CIRCLE & \LEFTcircle \space (CAs must maintain RT logs) \\
& \textbf{Log as revocation status responder} & \Circle & \Circle & \CIRCLE \\
& \textbf{Deployment cost} & Low (existing CT logs) & Medium (a modification of CT logs) & High (deployment of new logs) \\
& \textbf{Revocation status delivery cost} & \APLnot{\APLdown} & \APLnot{\APLdown}  & Low \\
\\
\midrule
& \textbf{Evaluation according to framework~\cite{chuat2019sok}}
\\
\midrule
\textit{A} & \textbf{Supports CA revocation} & \CIRCLE & \CIRCLE & \CIRCLE \\
& \textbf{Supports damage-free CA revocation} & \CIRCLE & \CIRCLE & \Circle \\
& \textbf{Supports leaf revocation} & \CIRCLE & \CIRCLE & \CIRCLE \\
& \textbf{Supports autonomous revocation} & \LEFTcircle & \LEFTcircle & Not specified \\
\midrule
\textit{B}  & \textbf{Supports delegation} & \CIRCLE & \CIRCLE & Not specified \\
& \textbf{Delegation w/o key sharing} & \CIRCLE & \CIRCLE & Not specified \\

\midrule
\textit{C}  & \textbf{Supports domain-based policies } &  \LEFTcircle &  \CIRCLE & \Circle \\
& \textbf{No trust-on-first-use required} & \CIRCLE & \CIRCLE & Not specified \\
& \textbf{Preserves user privacy} & \CIRCLE & \CIRCLE & \Circle \\

\midrule
\textit{D} & \textbf{Does not increase page-load delay} & \CIRCLE & \CIRCLE & \CIRCLE \\
& \textbf{Low burden on CAs} & \CIRCLE & \Circle & \Circle \\
& \textbf{Reasonable logging overhead} & \CIRCLE & \LEFTcircle & \CIRCLE \\

\midrule
\textit{E}  & \textbf{Non-proprietary} & \CIRCLE & \CIRCLE & \LEFTcircle \\
& \textbf{No special hardware required} & \CIRCLE & \CIRCLE & \CIRCLE \\
& \textbf{No extra CA involvement} & \Circle & \Circle & \Circle \\
& \textbf{No browser-vendor involvement} & \CIRCLE & \CIRCLE & \Circle \\
& \textbf{Server compatible} & \CIRCLE & \CIRCLE & \Circle \\
& \textbf{Browser compatible} & \CIRCLE & \CIRCLE & \Circle \\

\midrule
\textit{F}  & \textbf{No out-of-band communication} & \Circle & \Circle & ? \\

\\ \multicolumn{5}{c}{\CIRCLE~--- true, \Circle~--- false, \LEFTcircle~--- explanation follows, \APLnot{\APLdown} --- cost of a selected underlying revocation status protocol}
\end{tabular}
}
\end{table*}

We evaluate the two types of postcertificates using the comprehensive framework developed by Chuat et al.~\cite{chuat2019sok}. The framework consists of grouped properties that 
delegation and revocation protocols can provide.
Table~\ref{tab:comparison} presents the results.

\subsubsection{Property group A} Both of our schemes allow revocation of CA certificates and leaf certificates through the submission of corresponding postcertificates. ``Damage-free CA revocation" property depends on the implementation of postcertificates. The property holds if a scheme can invalidate descendants of a CA issued after a specified invalidation date. Self-signed postcertificates are more flexible w.r.t. to this property since a self-signed postcertificate is generated by the private key owner who can include the revocation reason and invalidation date at the time of postcertificate generation. Potentially, this can express in a large number of self-signed postcertificates. However, CA-issued postcertificates are issued and distributed to the private key owners in advance or upon request, i.e., the number of issued postcertificates can be limited by the CAs.
A CA can provide ``Damage-free CA revocation" property by issuing postcertificates with particular reason codes and invalidation dates on-demand.

According to the evaluation framework, one of the most important properties of a revocation protocol is ``Autonomous revocation", which means that domain owners can ``decide on the validity period or revocation status of their own certificates" and ``without reliance on a CA, browser vendor, or log". We argue that the postcertificate schemes are also {\it autonomous by policy} since the certificate owner can provably initiate the revocation process of a certificate through 
several
third-party logs.
Submission of a postcertificate makes the revocation request public and transparent.
Clients do not need to contact the issuers to revoke. The issuing CAs must revoke within 
a
guaranteed time limit after the submission of a postcertificate.

\subsubsection{Property group B} The postcertificate schemes enable the delegation of the capability to revoke a certificate through the distribution of the corresponding postcertificates without sharing the private key of the certificate. 

\subsubsection{Property group C}
Postcertificates do not support domain-based policies by default, but such policies can be implemented as additional extensions to the postcertificates, similarly to the reason codes and invalidation dates. Postcertificates do not require trust-on-first-use. Self-signed postcertificates are signed with the private key of the corresponding certificate, while CA-issued postcertificates are signed with the private key of the issuing CA.
Postcertificates preserve user privacy in the sense that browsers do not need third-party communication to validate a postcertificate. Web clients do not use postcertificates nor CT logs for revocation status checks. The security and privacy of revocation checks depend on the underlying revocation status protocol.

\subsubsection{Property group D} Postcertificates do not increase page-load delay, as they are not involved in connection establishment. We argue that ``CA-issued postcertificates" have a low burden on CAs since (1) the CAs can produce and deliver postcertificates to the clients during the certificate issuance process and (2) do not require any changes to 
the CT logs.

The logging overhead depends on the implementation of postcertificates and the revocation practices of the clients. Self-signed postcertificates can be issued by misbehaving clients in arbitrary quantities. Logs can prevent unlimited submission with additional insertion constraints to the logs. On the other hand, the CAs fully control the quantity of CA-issued postcertificates that can be submitted to the logs.

\subsubsection{Property group E} Postcertificates are non-proprietary and are easily standardizable. Postcertificates require no special hardware. Postcertificates can be issued through the same issuance procedure and at the same time as the corresponding certificates. CAs are involved in the issuance of (CA-issued) postcertificates. Moreover, CAs are obligated to monitor CT logs and initiate the revocation process for newly published postcertificates. No changes to web servers are necessary. The owner of a certificate revokes it simply by submitting  
the corresponding postcertificate
to a CT log. Similarly, no changes to web browsers are required since postcertificates are not directly used in the browsers.  

\subsubsection{Property group F} The framework specifies a benefit called ``No out-of-band communication". Revocation using postcertificates does require out-of-band communication with a chosen third-party log. However, for postcertificate revocation, communication with third-party logs is beneficial. A client can revoke their certificate without directly contacting the issuer, 
and
the existence of several logs provides 
reliability.

\subsection{Technical considerations}

The actual revocation time depends on the log-monitoring performance of a CA. The revocation time can be made independent of the postcertificate submission-to-publication delay. To achieve that, CAs must accept SCTs with the corresponding postcertificates ``out-of-band" and proceed with the revocations before the actual publication of the postcertificates.

Instead of monitoring all trusted logs, a CA may be required to monitor only some of them (e.g., logs that have their SCTs embedded in the original certificates issued by the CA). 

We expect that most major CAs are monitoring the trusted logs already since the CAs need to respond to any potential misissuance. Hence, we argue that the requirement for all CAs to monitor CT logs is reasonable, as it provides an opportunity to implement revocation transparency and improve oversight on the CT ecosystem. 

Direct revocation status checking via the CT logs is prohibitively expensive to implement in browsers. However, this proposal does not intend to turn CT logs into revocation status endpoints. The deployment of postcertificates would allow for the preservation of revocation requests, enable monitoring of revocations, and enforce autonomous revocation by policy. 

Any inconsistency, such as when a client submits several contradictory postcertificates, can be resolved using the timestamped entries of CT logs. We leave the priority of postcertificates and resolution of contradictions to the CA/Browser Forum. For example, the earliest/latest/any published postcertificate could provide a canonical revocation reason and/or invalidation date. It might also be mandated that a certificate is revoked immediately, even if the submitted postcertificate has an invalidation date in the future. We cannot predict how the current revocation practices~\cite{korzhitskii2021revocation} will change if the clients get the right to revoke at any time and for free. Hence, in Section~\ref{section:deployability} we model a case when corresponding postcertificates are submitted directly after certificate issuance. 

\begin{figure*}[t]
  \centering
  \includegraphics[trim = 0mm 250mm 0mm 260mm, clip, width=0.89\textwidth]{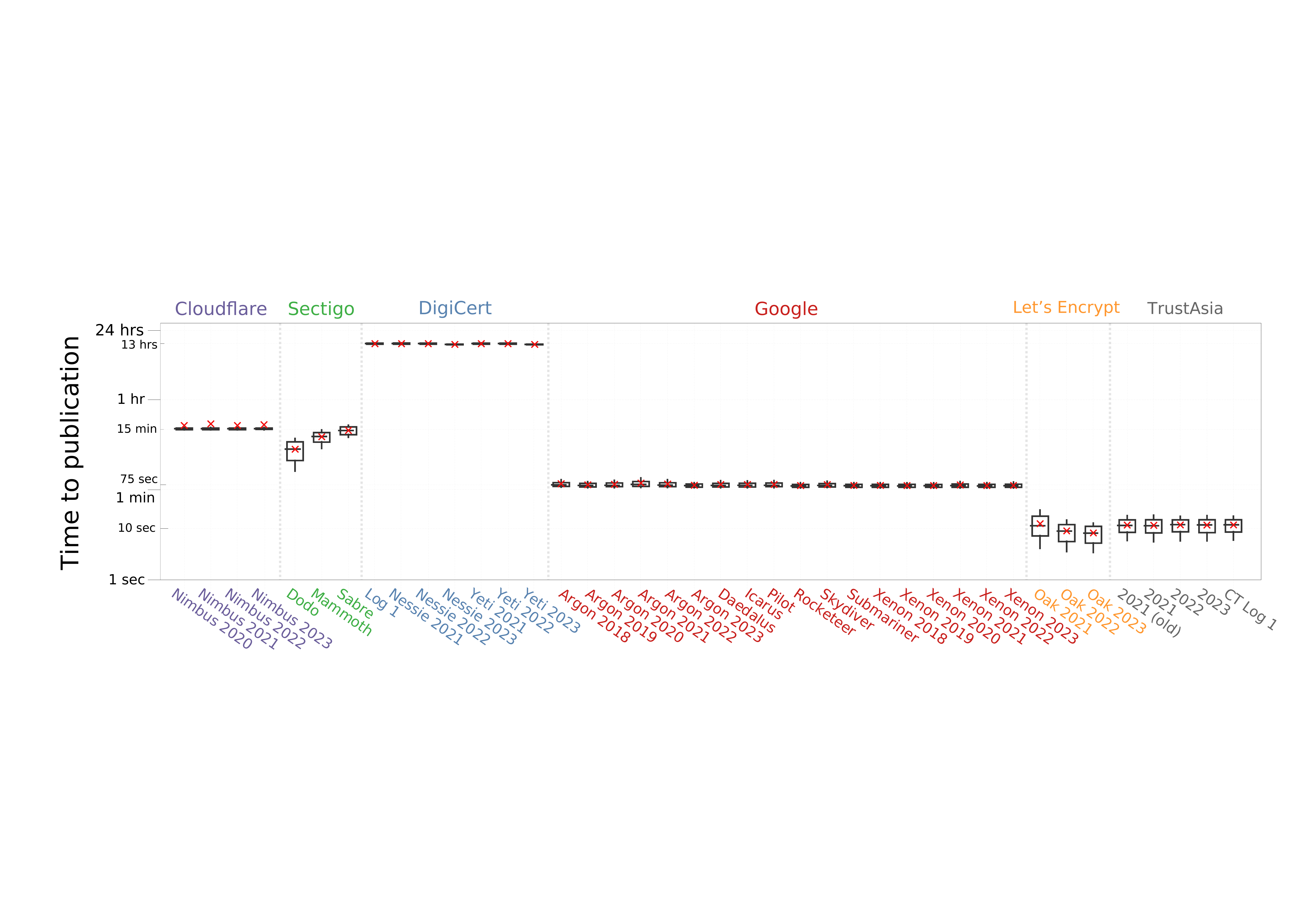}
\caption{Current submission-to-publication delay of CT logs.}
\vspace{-6pt}
  \label{fig:publication-delay}
  \vspace{-6pt}
\end{figure*}

\section{Performance of CT logs}
\label{section:performance}

We measure the performance of the existing CT logs to understand whether the logs can support the additional load that would be imposed on the logs by postcertificates. The measurement consisted of five phases:

\begin{enumerate}
    \item {\it Periodic sampling of Signed Tree Heads (STHs)} was used to determine 
    the growth rates of CT logs and the frequencies at which the logs sign STHs. An STH of a CT log contains the log's current size, a hash of the log's state, a timestamp, and a signature. In practice, some CT logs do not provide the STHs in a timely, consistent manner, e.g., due to caching. We attempt to circumvent caching by appending random and non-standard HTTP parameters to the requests.
    \item 
    {\it Periodic log size probing} was used to determine the actual number of available entries in the logs. We performed a binary search using the {\it get-entries} CT method to probe for the highest available entry number. 

    The probing is necessary because some CT logs publish entries asynchronously from the publication of a corresponding STH.
    \item We performed {\it certificate submissions} 
    to measure the submission-to-publication delay with respect to a reference clock. 
    Here, we extracted random entries from random CT logs and resubmitted each of them to every other CT log. 
    In case of a successful submission, the obtained SCT will contain a recent timestamp from the log's clock. However, the clocks between CT logs are not synchronized precisely. Thus, we timestamp all submissions and probes using a reference clock.
    \item {\it Log monitoring and entry number polling} was performed via {\it crt.sh}~\cite{crt.sh} to determine the final entry numbers of all our submissions. We use the entry numbers to infer the publication time of each submitted certificate relative to the STHs and probed log sizes. The above service ({\it crt.sh}) does not monitor some of the non-trusted logs: ct.browser.360.cn, Google Crucible/Solera, SHECA CT log, and Let's Encrypt Testflume. We exclude these logs from the analysis. 
    \item {\it Entry collision detection} allows us to find entries that are included in a log several times. Some CT logs incorporate identical certificate entries more than once. 
    The standard does not forbid this, stating that ``If the log has previously seen the certificate, it MAY return the same SCT as it returned before"\cite{RFCCT}. Furthermore, Chrome's CT policy explicitly treats submissions of the already-incorporated certificates identically to new submissions~\cite{ChromiumCTPolicy}. 
    We exclude such colliding entries from the analysis.
    
\end{enumerate}

\newcommand\varAvailableLogs{\var{45}}
\newcommand\varCertificatesSuccessfullySubmitted{\var{16K}}
\newcommand\varCollisionCertificates{\var{103}}
\newcommand\varMaxDelaySubmissionPublicationReferenceClockHour{\var{13.0 hours}}
\newcommand\varMeasurementEnd{\var{2021-06-22}}
\newcommand\varMeasurementStart{\var{2021-06-15}}
\newcommand\varMedianDelaySubmissionPublicationReferenceClockMin{\var{6.4 minutes}}
\newcommand\varMinDelaySubmissionPublicationReferenceClockSec{\var{1.0 seconds}}
\newcommand\varProbesPerLogPerHour{\var{348}}
\newcommand\varSTHPerLogPerHour{\var{311}}
\newcommand\varSlowestAverageSTHRate{\var{1.0}}
\newcommand\varStaticLogs{\var{5}}
\newcommand\varSuccessfulSubmissionsPerLogPerHour{\var{15}}
\newcommand\varSuccessfulSubmissions{\var{108K}}

\subsection{Summary}
The performance measurement of \varAvailableLogs{} available logs started on \varMeasurementStart{} and ended on \varMeasurementEnd{}. 
During this period, we performed \varSuccessfulSubmissions{} successful submissions of \varCertificatesSuccessfullySubmitted{} certificates, which is on average \varSuccessfulSubmissionsPerLogPerHour{} submissions to a log per hour; \varCollisionCertificates{} certificates collided, i.e., have been included in a CT log more than once.
Most of the logs were active during the measurement, i.e., they increased in size. 
However, \varStaticLogs{} older Chrome-untrusted (but available) CT logs remained frozen, even under our active submission attempts. 
Across our submissions, the minimum, median, and maximum delays from submission to publication are \varMinDelaySubmissionPublicationReferenceClockSec{}, \varMedianDelaySubmissionPublicationReferenceClockMin{}, and \varMaxDelaySubmissionPublicationReferenceClockHour{} respectively. 
On average, we have been performing \varProbesPerLogPerHour{} log size probes in each available log per hour and sending \varSTHPerLogPerHour{} STH requests to each available CT log per hour. However, the frequencies at which CT logs sign new tree heads vary drastically.
The lowest average STH signature rate among active CT logs is \varSlowestAverageSTHRate{} STHs per hour. We cannot precisely determine the fastest log update rate due to the chosen granularity of our measurement (i.e., probes roughly every 10 seconds). However, through closer analysis we found that 
some log implementations (e.g., Google’s and Let’s Encrypt’s logs) publish new tree heads at a sub-second rate. The following subsections provide CT log performance metrics most relevant to the deployment of postcertificates. Additional metrics can be found in the Appendix.

\subsection{Submission-to-publication delay}\label{sec:sub-to-pub}
We specify the submission-to-publication delay as {\it the delay between a submission request and an earliest STH response that covers the submission with respect to a reference clock} (see Section~\ref{section:performance}). Figure~\ref{fig:publication-delay} presents the measured delays for every log. 
For each log, we show the 10-percentile (bottom marker), 25-percentile (box bottom), median (middle marker), 75-percentile (box top), 
90-percentile (top marker), and the average value (a cross).

\begin{figure}[t]
  \centering
  \includegraphics[trim = 0mm 4mm 0mm 4mm, width=0.47\textwidth]{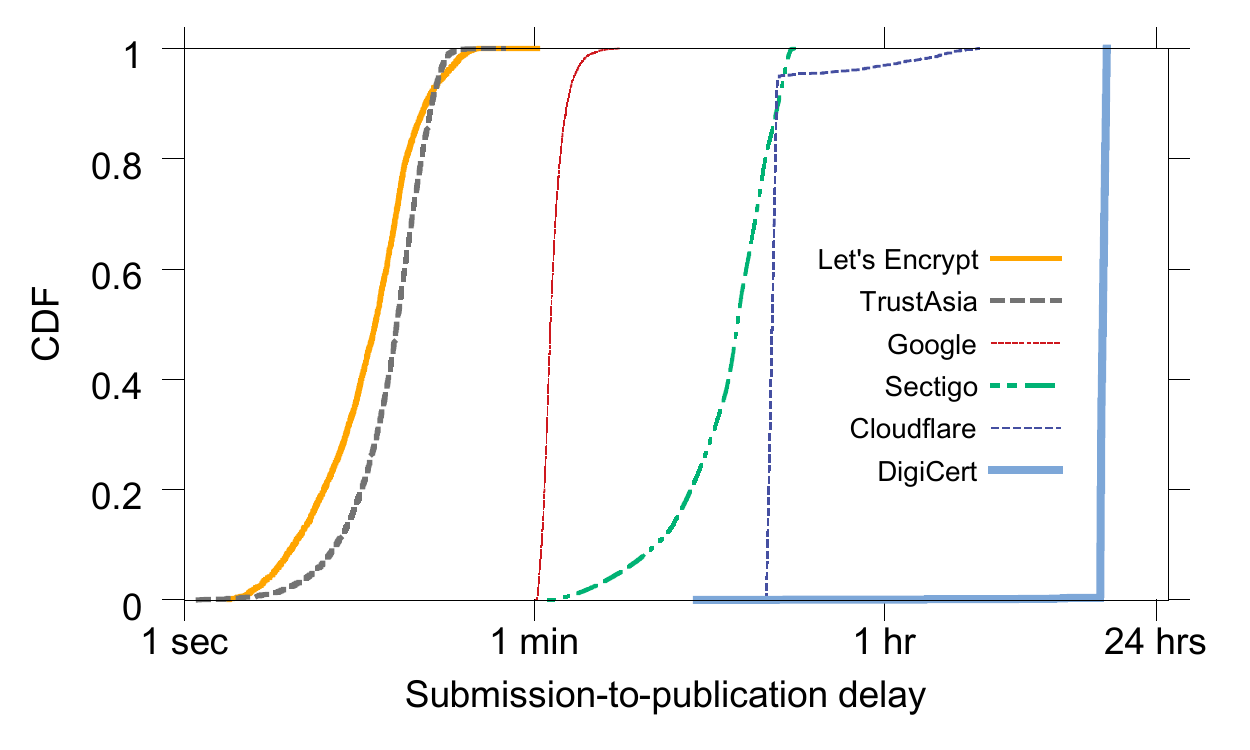}
  \vspace{-6pt}
\caption{CDFs of submission-to-publication delay per log operator.}
  \label{fig:submission-delay-cdf}
  \vspace{-6pt}
\end{figure}

\begin{figure*}[t]
  \centering
   \hspace*{-7pt}\includegraphics[trim = 0mm 0mm 0mm 0mm,width=0.84\textwidth]{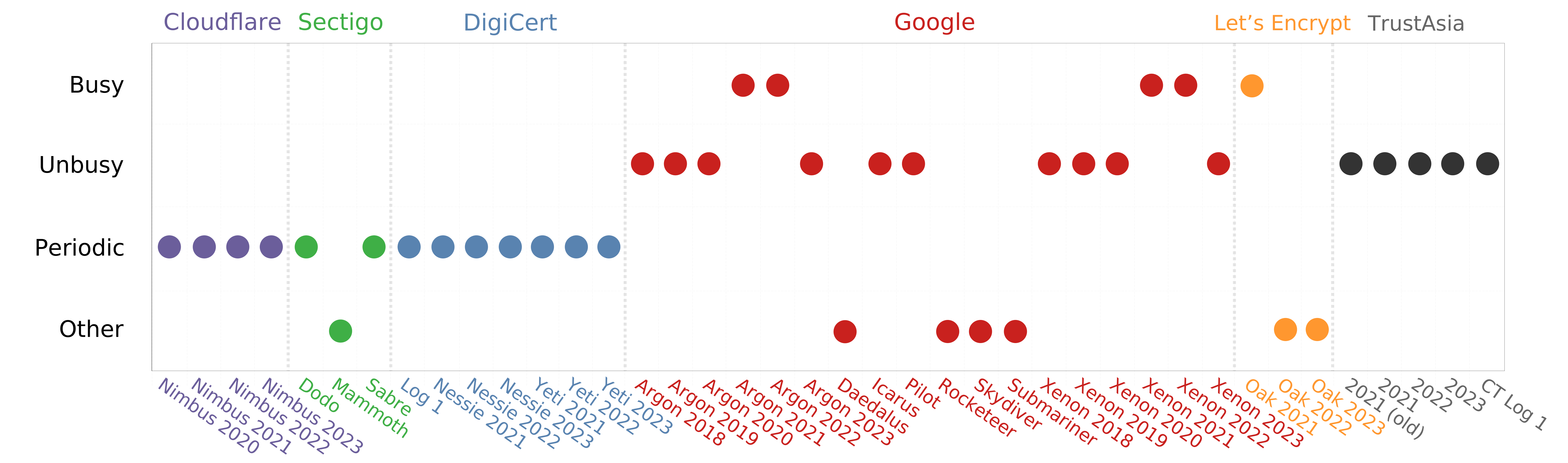}
    \vspace{-6pt}
 \caption{Classification of logs by their update behavior.}
 \label{fig:log-classes}
  \vspace{-0pt}
\end{figure*}

The fastest logs (Let's Encrypt, TrustAsia) publish certificates in a few seconds. Google publish certificates approximately in a minute. Most logs publish within an hour, while DigiCert publish submissions with an almost 13-hour delay. Figure~\ref{fig:submission-delay-cdf} shows the overall
distributions of submission-to-publication delays summarized per log operator.


\subsection{Log updates}

\newcommand\logClassOne{\var{Busy}}
\newcommand\logClassTwo{\var{Unbusy}}
\newcommand\logClassThree{\var{Periodic}}
\newcommand\logClassOther{\var{Other}}

Using several simple quantitative metrics based on STH timestamps (log's local time) and STH request times (our reference clock time), we identified three main classes of CT logs (see Figure~\ref{fig:log-classes}).
The first class of logs (\logClassOne) has a very high STH update frequency. For these logs, the granularity of our measurement does not allow us to precisely determine the STH update frequencies. We place a CT log to this class whenever more than 90\% of the successful submission requests are followed by an updated STH.
The second class of logs (\logClassTwo) includes CT logs with rare submissions that almost always grow by one entry whenever a log update happens. Note, that this growth typically occurs due to our active submission process and/or performance monitoring by Google. We place a log to this class whenever at least 90\% of its STH updates increment the log's size by one.
The third class of logs (\logClassThree) includes CT logs that have STHs timestamped and signed at a regular interval, or in the case of Cloudflare, some multiple of a fixed value. In all Cloudflare logs, the update intervals are of a multiple of 120 seconds. The Cloudflare Nimbus 2021/2022 logs almost always operate with a 120-second update interval, while the other less loaded Nimbus logs increase update interval by two or three times.
Sectigo Dodo and Sabre provide STH timestamps with an interval of $\approx 600$ seconds. DigiCert Nessie/Yeti 2023 logs provide STH timestamps that are slightly further than 600 seconds apart, while the other logs by DigiCert provide STH timestamps at intervals that are slightly further than 3,600 seconds apart. In Cloudflare 2021/2022, DigiCert 2021/2022 logs, Sectigo Dodo, and Sabre, we observed little variance in the time between updates.

\begin{figure}[t]
  \hspace*{-4pt}\includegraphics[trim=0mm 15mm 0mm 2mm,width=0.525\textwidth]{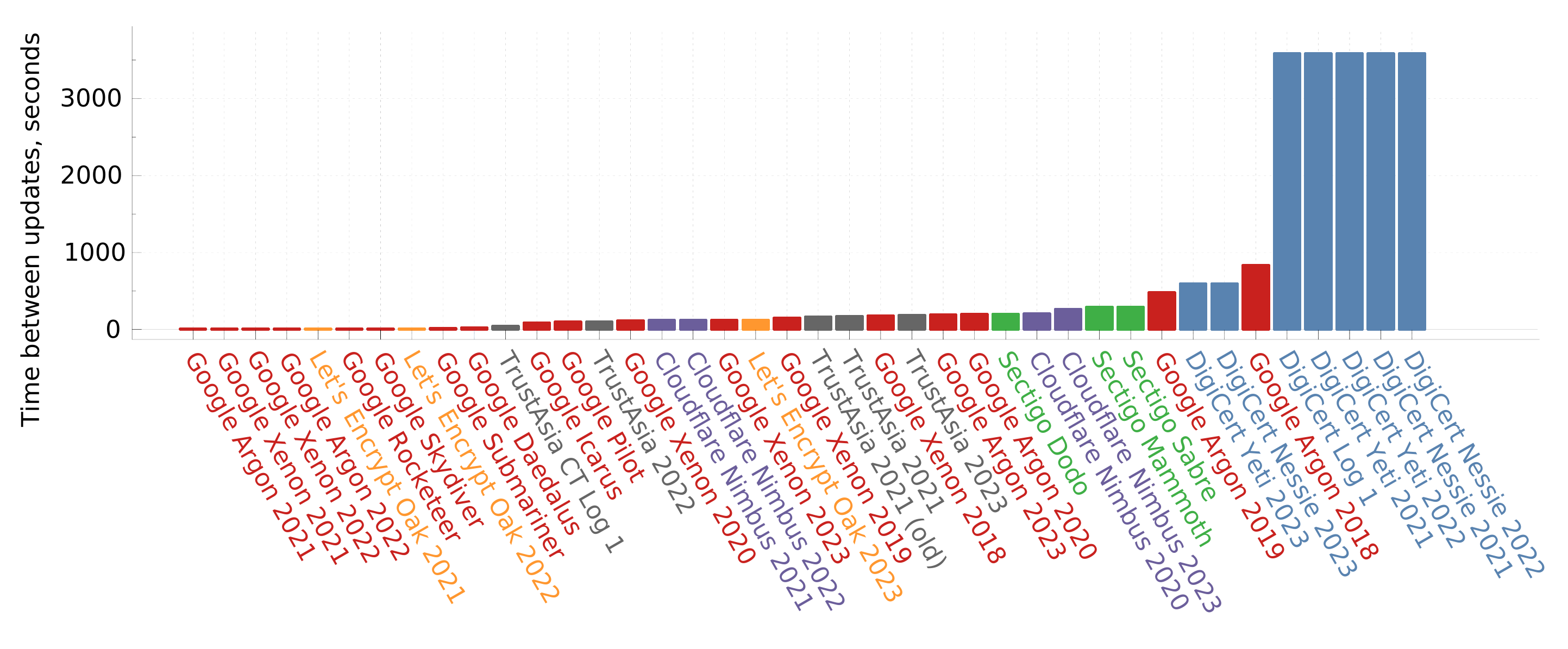}
  \vspace{-6pt}
  \caption{Upper bound on the average delay between consequent log updates.}
  \label{fig:delaysth}
\end{figure}

Some of the logs do not belong to any of the above classes. For instance, Sectigo Mammoth, similarly to other CT logs by Sectigo, often, but not always has 600-second intervals between the updates. Let's Encrypt Oak 2022/2023 neither seem to have a predetermined STH update period nor implement a particular threshold on the number of submissions to trigger an update. While Let's Encrypt Oak 2021 was highly loaded and almost always had a new STH to deliver upon a request, this was not the case for Oak 2022 and 2023. In these logs, we observed STH timestamp intervals of 1-to-198 seconds and 1-to-605 seconds, respectively. Note, that we miss many STHs issued with frequencies higher than our measurement's request frequency, i.e., STHs issued more frequently than every 10 seconds (due to our probe granularity). Both logs have significant variations in the number of entries incorporated during each update. We observed even higher variations in STH update intervals for some Google logs (e.g., Daedalus, Rocketeer, Skydiver, Submariner). CT logs by Google and Let's Encrypt are very fast at updating STHs, while longer update intervals are typically associated with periods of low submission rates. To plot the upper bound on the average delays between log updates (see Figure~\ref{fig:delaysth}), we look at consequent updates of STHs with respect to our clock. Note, that log STH updates can be somewhat independent of actual certificate publication. For example, DigiCert logs update their STHs every hour. However, the submission-to-publication delay is 13 hours.

\begin{figure}[t]
  \hspace*{-0pt}\includegraphics[trim = 4mm 18.5mm 13.5mm 3mm,clip,width=0.51\textwidth]{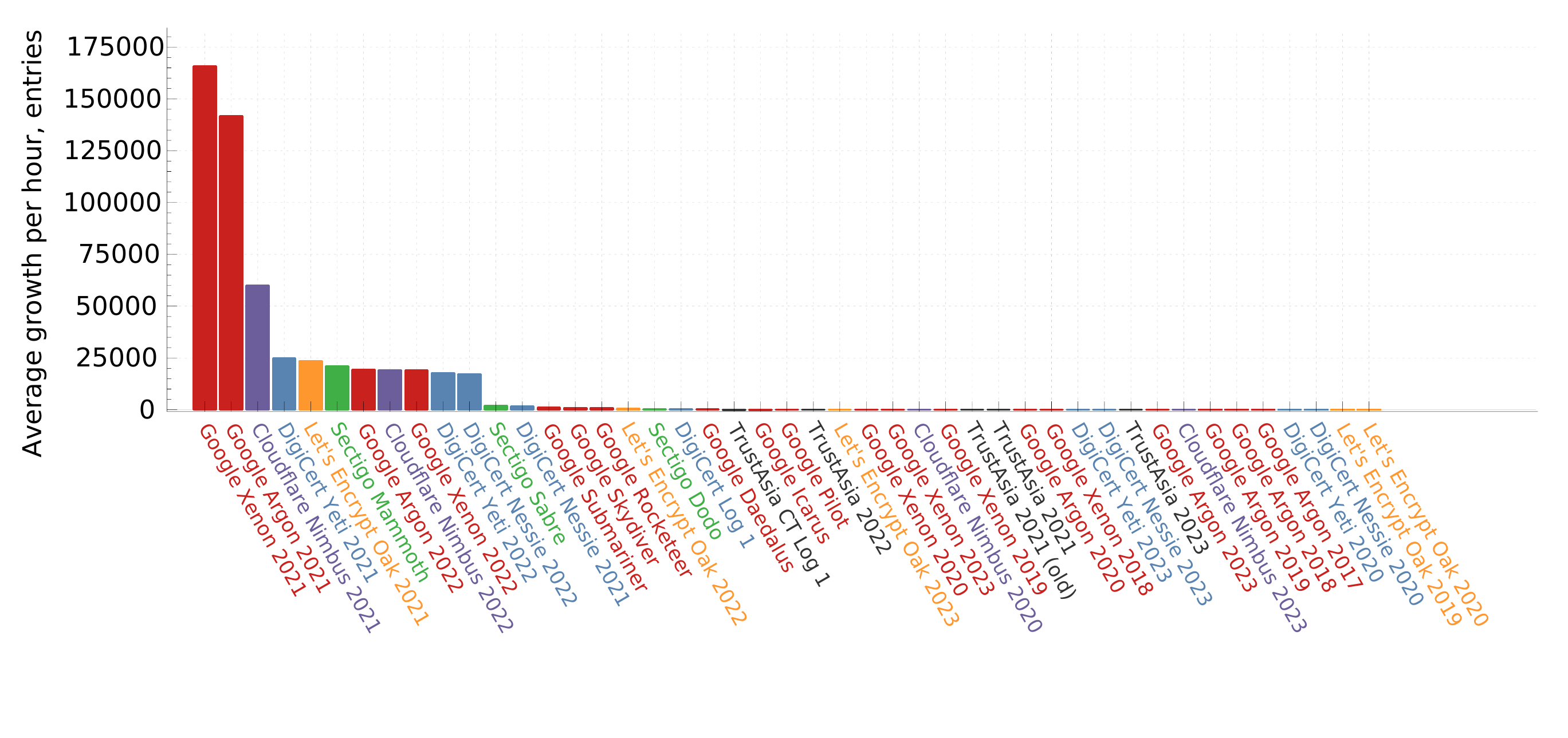}
   \vspace{-15pt}
  \caption{Average hourly growth rate of the logs.}
  \label{fig:growthRate}
\end{figure}

Figure~\ref{fig:growthRate} shows the growth rates of CT logs during our study. The average hourly growth rate of Google Xenon 2021 and Argon 2021 combined surpasses the total growth rate of all of the rest of the logs. Only 11 CT logs are above 17K submissions per hour, while the rest are below 2K submissions per hour. It demonstrates the dominant role of Google in the CT infrastructure. We argue that the infrastructure would benefit from a more even and reliable load distribution. In the case of a failure of Google's logs, the certificates from across the Internet must not overwhelm the rest of the operators.

\subsection{Entry collisions}
During the 
study,
\varCollisionCertificates{} of our submitted certificates 
collided with some external submissions. 
While most logs provide old inclusion timestamps instead of incorporating a certificate again, some identical entries have been included in Cloudflare and DigiCert logs several times, but with different timestamps.

\begin{figure}[t]
  \centering
  \includegraphics[trim = 6mm 2mm 15mm 0mm,clip, width=0.48\textwidth]{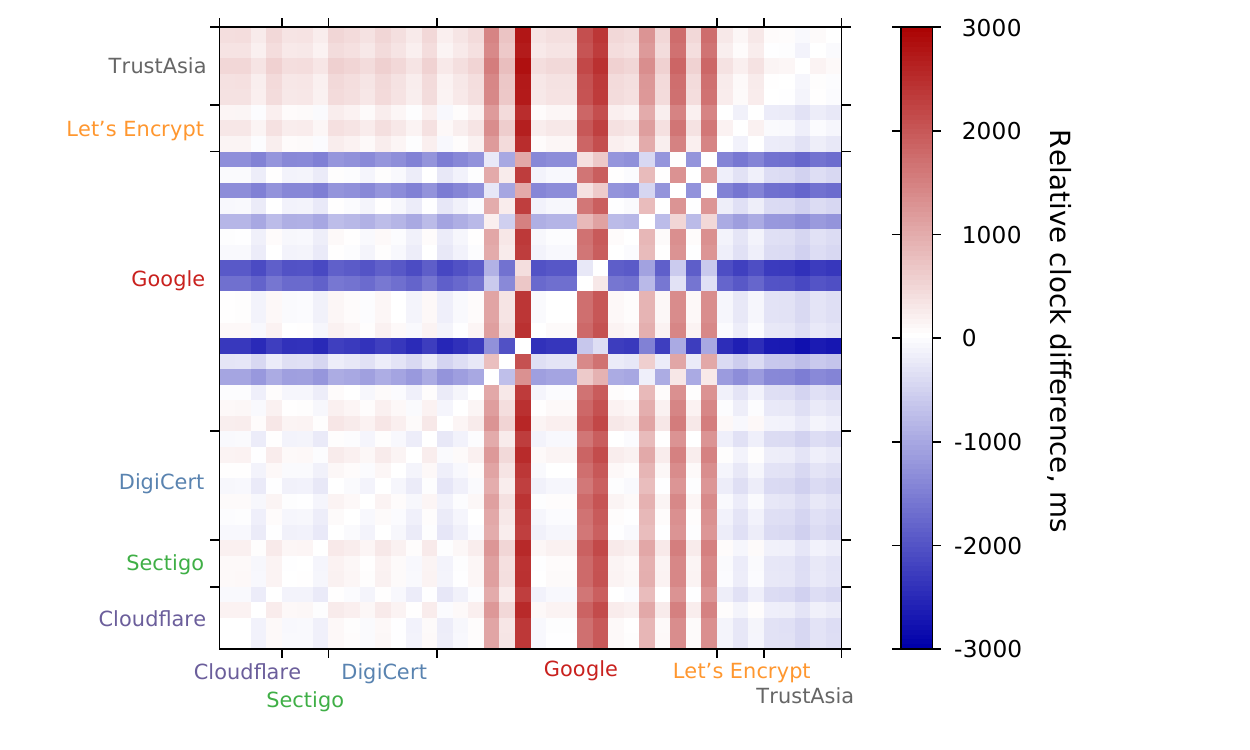}
 \vspace{-8pt}
\caption{Relative pairwise log clock difference. Logs are ordered similarly to the previous figures (e.g., Figure~\ref{fig:log-classes}).}
  \label{fig:sync}
  \vspace{-0pt}
\end{figure}

\subsection{Log synchronization} 
\label{subsection:sync}
CT logs provide an essential timestamping service for CAs. In our proposal, CAs sign certificates as revoked with timestamps strictly after the time provided by the earliest SCT timestamp obtained from a postcertificate submission.
Hence, it is important that the logs are well-synchronized. 

To understand the level of synchronization already achieved by CT logs, we probed the clocks of CT logs via the timestamps provided in SCTs of successful submissions. 
Using this data, we compare the timestamps to our reference clock and calculate relative pairwise clock offset between CT logs. 
For these calculations, 
we first calculate the median offset relative to our clock and then compare the relative offsets for each pair of logs. 
The results are shown in Figure~\ref{fig:sync}. 

All logs are somewhat synchronized and appear to have a median relative pairwise clock offset of at most 3 seconds. Surprisingly, some of the CT logs with the highest relative differences are Google logs. The clock offsets of all CT logs are much smaller than their $MaxMergeDelay$s and the submission-to-publication delays. Thus, we argue that the current level of synchronization is sufficient to implement revocation via postcertificates, as long as the CAs' signed revocation statuses contain status timestamps that are past the time of the earliest SCT issued for a postcertificate.


\section{Deployability}
\label{section:deployability}

\subsection{Additional load on logs}
\begin{figure}[t]
  \centering
  \includegraphics[trim = 8.5mm 196.5mm 30mm 0mm,clip,width=0.47\textwidth]{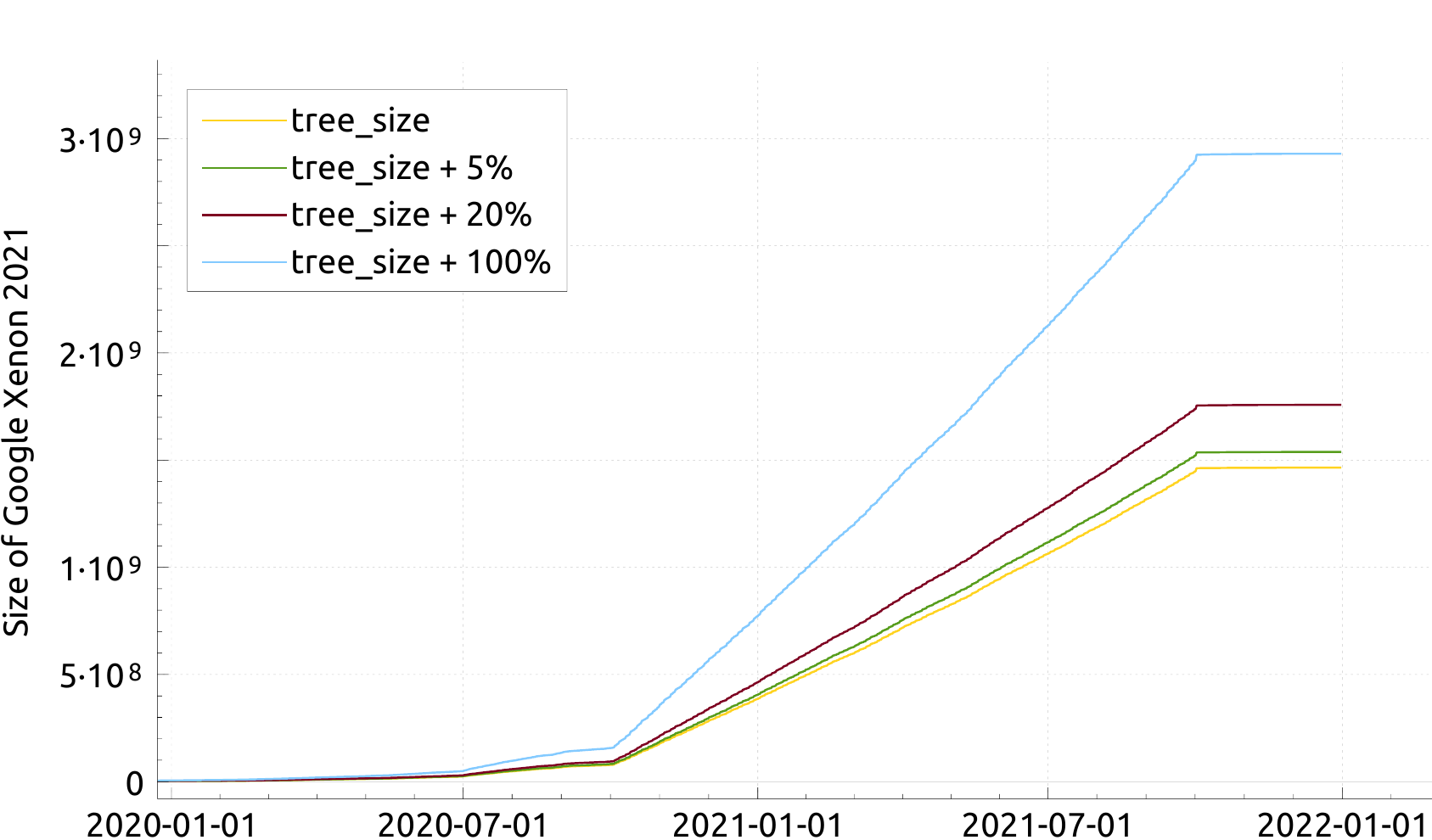}
  \caption{Historical growth of Google Xenon 2021 with added 5, 20, 100\% of corresponding postcertificates.}
    \label{fig:model}
\end{figure}

We model the potential impact of CA-issued postcertificates on the historical growth rate of Google Xenon 2021 --- the fastest-growing log during our study.
Figure~\ref{fig:model} demonstrates the historical growth rate of the log, along with several scenarios. Assuming that each entry in the history of the log is a unique certificate that requires the issuance of a single postcertificate, we add 5, 20, or 100\% of additional postcertificates that are submitted to the log directly after issuance (worst case). We assess the additional submission load to be well within the capabilities of the existing logs.

\subsection{Revocation process}
\label{section:revocation-delay}

\subsubsection{Revocation request}
While the current CA/Browser forum requirements~\cite{CABR20} do not require CAs to monitor CT logs, some already do~\cite{monitors}.
For an ``in writing" revocation request, a CA is obligated to revoke the certificate within 24 hours. CAs must maintain a ``24x7 ability to accept and respond to revocation requests". 
However, the existing means for delivery of revocation requests/notifications, such as online websites, postal services, email, and phone do not provide cryptographic proofs of delivery, nor are they
sufficiently reliable, compared to CT submission requests. For notifications that have not been received in writing, the CA must revoke a certificate within 5 days. The time before an updated status is published may vary, depending on a revocation reason and the party that initiated the revocation. A list of parties capable of starting a revocation includes a Subscriber (i.e., client), an Issuing CA, and Registration Authorities. Other third parties can also report problems related to certificates. In the postcertificate schemes, the revocation process starts with the submission of a postcertificate to a CT log.

\subsubsection{Status update} 
\label{sub:status-update}
The proposed postcertificate schemes do not affect the delivery of statuses to the clients by an underlying revocation status protocol.
In the current WebPKI, although CAs can change their OCSP revocation statuses in an instant, the shortest allowed status validity period is still very long, and the CAs are permitted to have long status update delays~\cite{CABR20}. In particular, the validity period of OCSP statuses is allowed to be between 8 hours and 10 days. If the validity period is longer than 16 hours, then the maximum time before an updated status must be published is limited by $min[(ValidityPeriod - 8 hours), 4 days]$. But if the validity period of a revocation status is shorter than 16 hours, then an updated status must be made available within $ValidityPeriod/2$. Note, that several contradicting statuses can be valid at the same time. 
The introduction of postcertificates is compatible with the above policies. Additionally, logged postcertificates provide a transparent record of revocation requests. 

\subsubsection{Revocation delay}
\label{sub:revocation-delay}
The current revocation delay ($T_{Revocation}^{Current}$) can be broken down into three consequent delays: the time that it takes to deliver the request ($T_{Request Delivery}^{Current}$), the processing time of the request ($T_{Request Processing}^{Current}$), and the time it takes to update the status 
($T_{Update}^{Status}$).
Thus, we have:
\begin{align}
   & T_{Revocation}^{Current} = T_{Request Delivery}^{Current} 
   + T_{Request Processing}^{Current} 
   + T_{Update}^{Status},
\end{align}
where 
$(T_{Request Processing}^{Current} + T_{Update}^{Status})$
must be less than or equal to 5 days.

Similarly, for a revocation performed using a postcertificate the revocation time ($T_{Revocation}^{Postcert}$) can be calculated as:
\begin{align}
    T_{Revocation}^{Postcert} = T_{Publication}^{Postcert} 
    + T_{MonDiscovery}^{Postcert} 
    + T_{Update}^{Status},
\end{align}
where $T_{Publication}^{Postcert}$ and $T_{MonDiscovery}^{Postcert}$ correspond to the time that it takes to publish the postcertificate in a CT log 
(i.e., the submission-to-publication delay)
and the time that it takes for the CAs monitor to discover the postcertificate, respectively.
Here, the publication time is upper bounded as $T_{Publication}^{Postcert} \leq Max Merge Delay$.  Furthermore, when the $Max Revocation Delay$ is defined relative to the publication time, then we require that
$T_{MonDiscovery}^{Postcert} + T_{Update}^{Status} \leq MRD_B$.  
Given these observations, the worst-case revocation delay in this case is upper bounded by: 
\begin{align}
T_{Revocation}^{Postcert} \leq MMD + MRD_B.
\end{align}
In the case that the $Max Revocation Delay$ is defined relative to the submission time,
$MRD_A$ becomes the upper bound; i.e, 
\begin{align}
T_{Revocation}^{Postcert} \leq  MRD_A.
\end{align}
As noted earlier, $MRD_A$ must be selected such that $MRD_A > MMD$, whereas $MRD_B$ can be any $MRD_B > 0$ sufficient to allow discovery and status update.

The time to discovery can be substantially improved for cases when clients and CAs cooperate. For this purpose,
we suggest that CAs should allow clients to submit the SCTs they obtain from the logs together with a revocation request (e.g., in the form of the postcertificate + SCTs) directly to the CA.  

Since the response times of logs (as shown in this paper) is short in most cases, the approach would allow responsible CAs to perform revocations requests even quicker. Given the postcertificate and electronic proof of the postcertificate submission and the time of the submission (i.e., an SCT), this submission process is expected to allow the same -- or faster -- timeline as the current revocation practices allows, even in cases when the client uses a log with large merge delays. Furthermore, revocation using postcertificates provides additional control to clients over the revocation status.

In practice, {\it MaxMergeDelay}
is 24 hours for CT logs~\cite{AppleCTLogProgram,ChromiumCTPolicy}, 
$T_{Update}^{Status}$
depends on the underlying revocation status protocol (currently, up to 4 days in OCSP)~\cite{CABR20}, and the {\em Maximum Revocation Delay} 
must
be decided by a standardizing body such as CA/Browser forum. The monitoring delay $T_{MonDiscovery}^{Postcert}$ depends on the performance of the logs and monitors.
We next take look closer at the performance of one such monitor.

\subsubsection{Monitoring delay}
\begin{figure}[t]
      \hspace*{0pt}\includegraphics[trim = 10mm 20mm 5mm 19mm,width=0.48\textwidth]{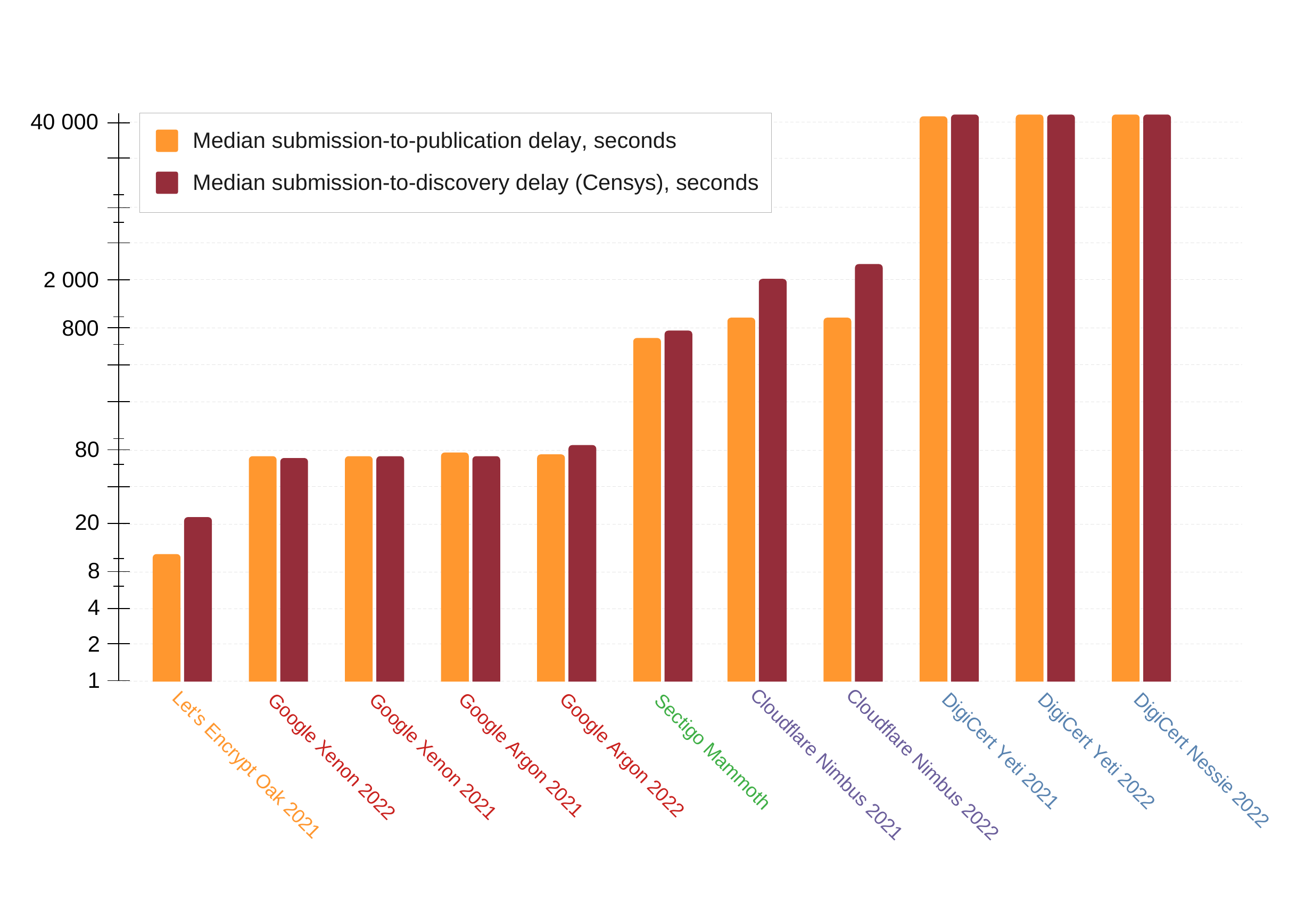}
       \vspace{-1pt}
      \caption{Measured median submission-to-publication delay v.s. median submission-to-discovery delay reported by Censys~\cite{censys15} on log scale.}
      \label{fig:censys}
    \end{figure}

We use data obtained from Censys~\cite{censys15} to compare the performance of their monitor to our submission-to-publication measure. For every found certificate, Censys provides SCT timestamps and monitor inclusion timestamps of the corresponding log entries.
In Figure~\ref{fig:censys} we present the median submission-to-discovery delay for the precertificates that Censys found during the measurement period in the 11 most actively used logs.

For most CT logs, Censys discovers and includes new certificates soon after the publication of the certificates in the logs. Moreover, for some Google logs, Censys manages to include certificates before we can detect their publication using STH probes. We assume that for these logs Censys must be sending requests at a frequency that is higher than ours. However, for Cloudflare Nimbus 2021/2022 logs, there is a considerable delay between the publication of a certificate and its inclusion by the monitor. Nevertheless, submission-to-publication and submission-to-discovery time are well within $MaxMergeDelay$ for all logs.
Overall, we conclude that publication and monitoring delays are sufficiently low for implementing postcertificate revocation.

\subsection{Incremental, voluntary, and test deployment.}
CA-issued postcertificates are incrementally deployable. For instance, CAs could (i) start with the issuance and logging of postcertificates upon revocation without postcertificate distribution, (ii) develop and deploy postcertificate monitoring routines, and (iii) start distributing postcertificates to clients for revocation. No modifications to CT logs are necessary for the deployment of the scheme.
Moreover, the deployment of the scheme can be voluntary. That is, in order to improve the transparency and trustworthiness of the issued certificates and corresponding revocation statuses, a CA may voluntarily implement postcertificates and commit to postcertificate monitoring. Similarly, the real-world deployment of CA-issued postcertificates can be tested by a CA without any changes to the CT logs and would only require changes to the testing CA.

\section{Related work}
\label{section:relatedwork}

Chuat et al.~\cite{chuat2019sok} provide a comprehensive evaluation framework 
for
revocation and delegation protocols. 
In this paper, we use the framework to evaluate the proposed postcertificate schemes.
Certificate Transparency~\cite{RFCCT} and Revocation Transparency~\cite{laurie2012revocation} were originally proposed by Google. The latter describes how to use 
Sparse Merkle Trees (SPTs)
and sorted lists for efficient storage and lookup of revocation statuses. However,  Ryan~\cite{ryan2014enhanced} reported that the proposal is not practical due to the large size of inclusion proofs. Dahlberg et al.~\cite{dahlberg2016efficient} provide a detailed overview of 
SPTs 
and their performance. 

Mueller et al.~\cite{mueller2019let} propose submission of certificates with specially formed names in CT logs to revoke OpenPGP keys. The proposal introduces a revocation service that receives a signed revocation request (revocation bytes) from a holder of a private key; the service issues a special certificate that contains revocation bytes and the name of the key that is to be revoked. The service then transfers the certificate to a CA; the latter signs the certificate and submits it to a CT log. To check the revocation status of an OpenPGP key a client must fetch all certificates from all CT logs. Thus, the logs act as responders delivering revocations to clients. 

In our proposal, CAs are required to monitor CT logs. Li et al.~\cite{li2019certificate} studied the reliability of existing CT monitors. 

Instead of classical revocation protocols like CRL~\cite{RFCPKI} and OCSP~\cite{RFCOCSP}, some software vendors use proprietary lists of revocations~\cite{CRLSets,crlite}. Several alternative revocation status protocols have been developed~\cite{chariton2016dcsp,chariton2017ccsp,compressedocsp,smith2020let}, as well as novel PKIs~\cite{basin2014arpki,kim2013accountable,szalachowski2016ritm,szalachowski2016pki,yu2016dtki}; however, all of these 
protocols and PKIs
require large changes to the infrastructure. Similarly, many other semi-centralized and decentralized PKIs, revocation protocols and logs have been proposed~\cite{ssdpki,BlockVoke,lightledger,dledger,kubilay2019certledger,leibowitzctng,sermpinis2021detract,decentralized-cosigning,distrpki}. Matsumoto et al.~\cite{incentives} propose a decentralized framework that incentivizes CAs to monitor for certificate misissuance through financial penalties.

Tomescu et al.~\cite{tomescu2019transparency} propose a transparency log system that minimizes proof sizes and bandwidth usage at the cost of the increased append times. Kales et al.~\cite{ctprivacy} study effects of CT on user privacy and implement privacy-preserving and efficient membership testing for CT logs, which can potentially enable the use of the logs for direct and secure postcertificate lookup by clients. 

Due to the lack of a standardized revocation transparency method, Let's encrypt and DigiCert announced their mass revocation events in arbitrary ways, by publishing datasets of the revoked certificates on their websites and public forums~\cite{massDig2,massDig1,LE-datasets2020}. In the case of the Let's Encrypt mass-revocation, not all initially announced revocations have actually been performed. Some of the previously revoked and/or expired CA certificates (e.g., DigiNotar) are still present in trust root stores of the CT logs~\cite{korzhitskii2020characterizing}. 

Internet revocations have been measured in several studies. Liu et al.~\cite{liu2015end} found that a high fraction of served certificates was revoked (8\%), while CRLSets~\cite{CRLSets} by Google was only covering 0.35\% of all revocations. 
Chung et al.~\cite{muststaple18} performed a measurement and concluded that OCSP responders were not sufficiently reliable to support {\it OCSP Must-staple} extension. 
Zhu et al.~\cite{zhu2016measuring} found 0.3\% of certificates to be 
revoked,
assessed the
OCSP latency to be ``quite good", and showed that 94\% of OCSP responses were served using CDNs. Smith et al.~\cite{smith2020let} found that in the absence of a mass-revocation event, the revocation rate on the Internet was 1.29\%. 
Korzhitskii et al.~\cite{korzhitskii2021revocation} checked certificate status one day before expiration and 
onward, with the measurement capturing a mass revocation event.
The revocation rate was 0.35\% for Let's Encrypt (2.4\% during the mass-revocation event) and 1.5\% for the other CAs.
Moreover, it was found that most revocations stop being advertised shortly after the expiration of a revoked certificate. Kim et al.~\cite{windowsRev} measured revocation effectiveness of the code-signing PKI; the latter preserves revocation statuses indefinitely.

\section{Conclusion}
\label{section:conclusion}

We introduced a revocation transparency protocol with postcertificates. Many parties on the Internet can benefit from the deployment of the protocol since postcertificates make revocation requests transparent, accountable, persistent, and available for study. The protocol allows certificate owners to initiate the revocation process autonomously. Postcertificates are compatible with the current and future revocation status protocols, including the mandatory OCSP protocol. 
Postcertificates do not require changes to CT logs and can be deployed incrementally, voluntarily, or as an experiment.
We measured the performance of CT logs and provided insights into the potential deployment of the protocol. We conclude that the performance of existing CT logs is sufficient for the adoption of postcertificates.

\bibliographystyle{splncs04}
\bibliography{main}


\begin{figure*}[t]
  \centering
  \hspace*{-17pt}\includegraphics[clip,trim = 0mm 19mm 0mm 20mm, width=0.858\textwidth]{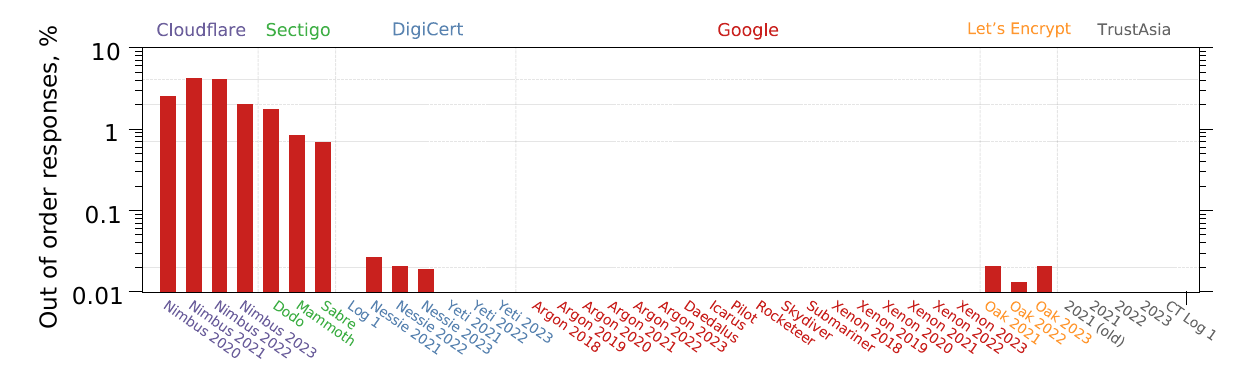}
  \vspace{-8pt}
\caption{Percentage of out-of-order STH responses for each log on a logarithmic scale.}
  \label{fig:outoforder}
  \vspace{6pt}
%
  \centering
  \hspace*{-15pt}\includegraphics[clip,trim = 0mm 0mm 0mm 0mm, width=0.85\textwidth]{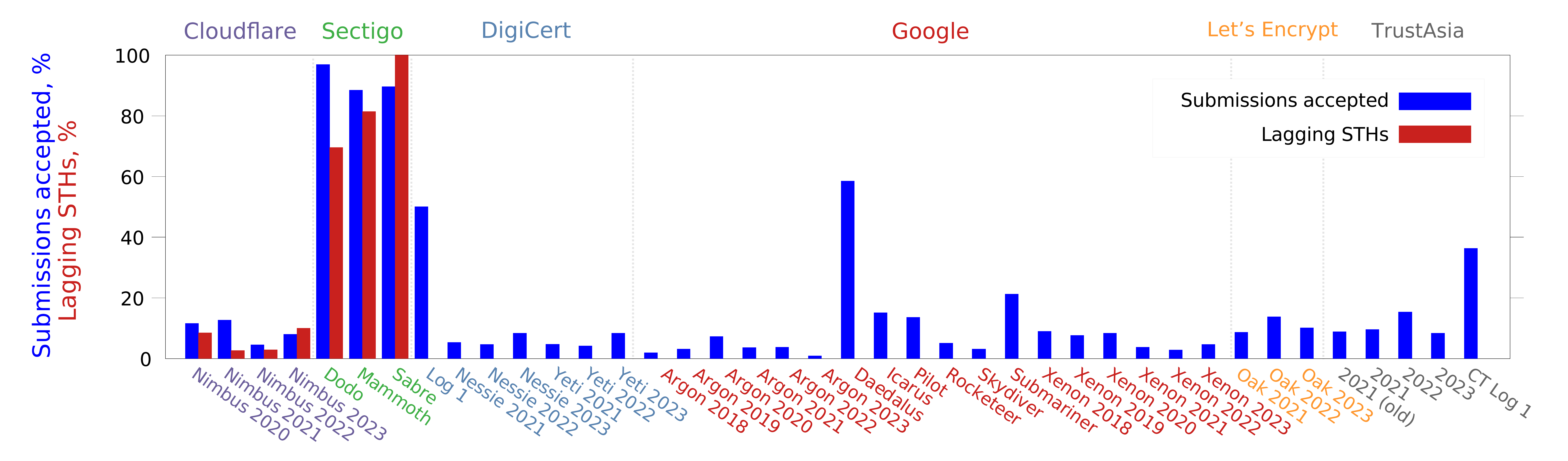}
\vspace{-8pt}
\caption{Percentage of random submissions accepted and percentage of lagging STHs.}
  \label{fig:sth-lag}
  \vspace{-0pt}
\vspace{6pt}
  \centering
  \hspace*{-17pt}\includegraphics[trim = 0mm 245mm 0mm 270mm, clip, width=0.865\textwidth]{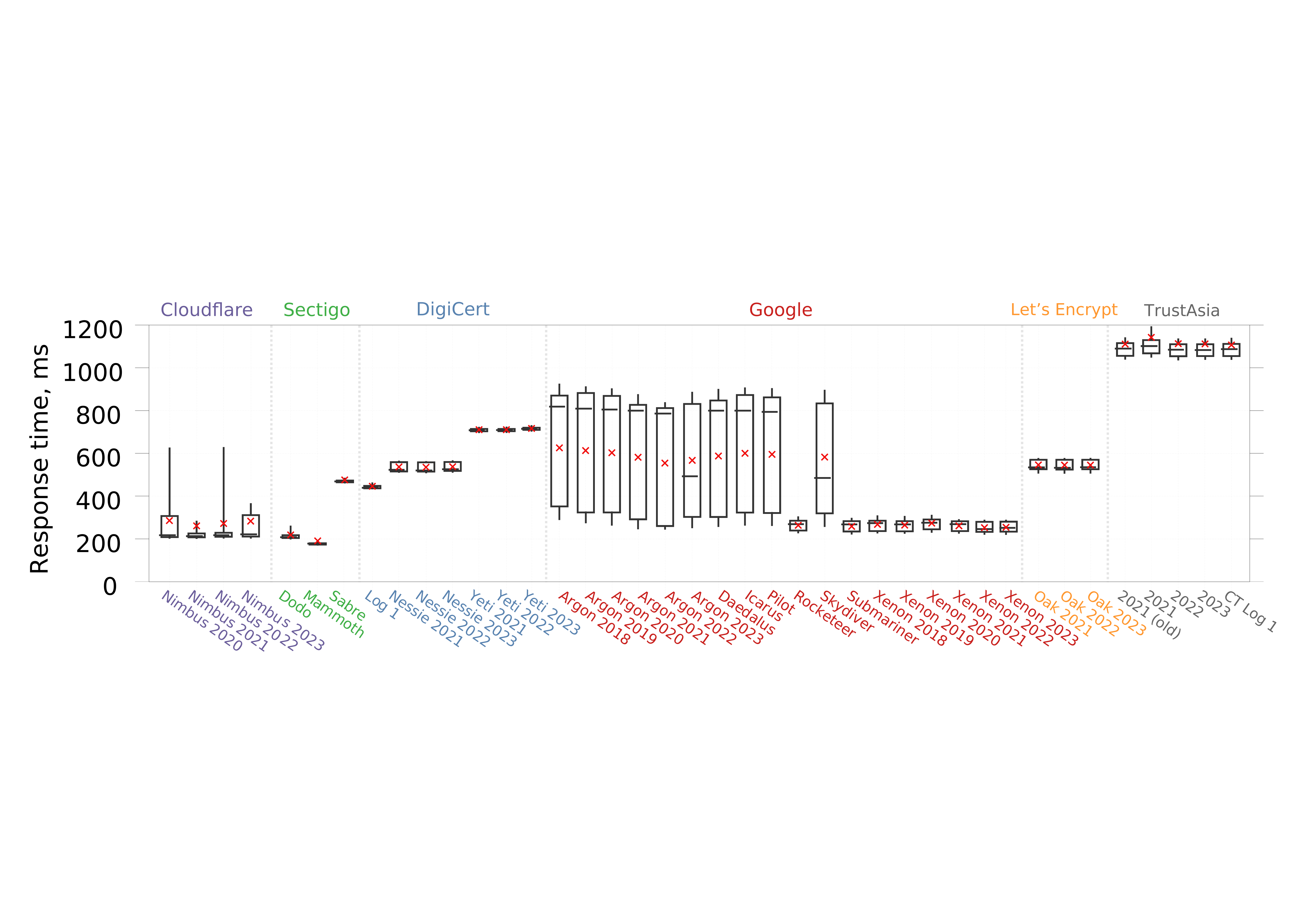}
\vspace{-8pt}
\caption{Submission request processing time.}
  \label{fig:submission-processing}
  \vspace{-0pt}
\end{figure*}

\balance 
\section*{Acknowledgment}
This work was supported by the Wallenberg~AI, Autonomous Systems and Software Program (WASP) funded by the Knut and Alice Wallenberg Foundation.

\section*{Appendix. Extra on CT log performance}
\label{appendix-a}

\subsection{Out-of-order log state responses}
We observed that only some logs provide out-of-order STH responses according to the timestamps/tree sizes. We suspect that this happens due to caching or load balancing. Figure~\ref{fig:outoforder} demonstrates the percentages of out-of-order responses per log on a logarithmic scale. 
All Cloudflare and Sectigo logs have a significantly higher percentage of out-of-order responses.
Such out-of-order responses can increase the time to log entry discovery for third parties.

\subsection{Lagging Signed Tree Heads}
In some logs, it is possible to access entries out of STH-advertised tree size bound. We specify an STH to be lagging if between an STH response with maximum advertised tree size $N$ and its predecessor it is possible to access an entry $M \geq N$. STHs could be lagging due to caching or load-balancing of STH responses. Figure~\ref{fig:sth-lag} demonstrates the percentages of lagging STHs along with the percentages of accepted submissions of randomly selected entries from other logs.
Note that only logs by Cloudflare and Sectigo produce lagging STHs. Lagging STHs increase the time to entry discovery of already available log entries for third parties.

\subsection{Submission request processing time}
For all successful certificate submissions, we calculate the delay between the start of a submission request and the end of the response. Similarly to Figure~\ref{fig:publication-delay}, Figure~\ref{fig:submission-processing} provides submission request processing delays for all logs. 

The observed request processing time across the logs is small enough to be considered negligible in the estimation of the submission-to-publication time and the overall revocation delays of the postcertificate schemes.

\end{document}